# The combinatorics of overlapping genes

Sophie Lèbre (1, 2, 3, 4) and Olivier Gascuel (4, 5)

(1) IMAG, UMR 5149 - CNRS & Université de Montpellier, France.
(2) CPBS, UMR 5236 - CNRS & Université de Montpellier, France.
(3) Université Paul Valéry Montpellier 3, France.
(4) Institut de Biologie Computationnelle, LIRMM, UMR 5506 - CNRS & Université de Montpellier, France.
(5) Unité Bioinformatique Evolutive, C3BI, USR 3756 - Institut Pasteur & CNRS, Paris, France.

**Abstract:** Overlapping genes exist in all domains of life and are much more abundant than expected upon their first discovery in the late 1970s. Assuming that the reference gene is read in frame +0, an overlapping gene can be encoded in two reading frames in the sense strand, denoted by +1 and +2, and in three reading frames in the opposite strand, denoted by -0, -1, and -2. This motivated numerous researchers to study the constraints induced by the genetic code on the various overlapping frames, mostly based on information theory. Our focus in this paper is on the constraints induced on two overlapping genes in terms of amino acids, as well as polypeptides. We show that simple linear constraints bind the amino-acid composition of two proteins encoded by overlapping genes. Novel constraints are revealed when polypeptides are considered, and not just single amino acids. For example, in double-coding sequences with an overlapping reading frame -2, each Tyrosine (denoted as Tyr or Y) in the overlapping frame overlaps a Tyrosine in the reference frame +0 (and reciprocally), whereas specific words (*e.g.* YY) never occur. We thus distinguish between null constraints (YY = 0 in frame -2) and non-null constraints (Y in frame +0 ⇔ Y in frame -2). Our equivalence-based constraints are symmetrical and thus enable the characterization of the joint composition of overlapping proteins. We describe several formal frameworks and a graph algorithm to characterize and compute these constraints. As expected, the degrees of freedom left by these constraints vary drastically among the different overlapping frames. Interestingly, the biological meaning of constraints induced on two overlapping proteins (hydropathy, forbidden di-peptides, expected overlap length …) is also specific to the reading frame. We study the combinatorics of these constraints for overlapping polypeptides of length $n$, pointing out that, (i) except for frame -2, non-null constraints are deduced from the amino-acid (length = 1) constraints and (ii) null constraints are deduced from the di-peptide (length = 2) constraints. These results yield support for understanding the mechanisms and evolution of overlapping genes, and for developing novel overlapping gene detection methods.

1. **Introduction**

In the late 1970s, it was discovered that a single DNA sequence may code for several *overlapping* genes, that is, protein-coding genes sharing a common section of DNA sequence read in different frames. The genetic code associates a section of DNA sequence (comprised of four nucleotides: *a, c, g, t*) with a protein, which is a sequence of amino acids. Each sequence of three nucleotides or *codon* codes for an amino acid (among 20, denoted as A, C, D, E, F, G, H, I, K, L, M, N, P, Q, R, S, T, V, W, Y; see Tab. 1, provided for completeness). Thus, shifting the reading frame by one or two nucleotides results in a different protein. As DNA consists of two strands, up to six reading frames are possible for a given DNA sequence. Several notations have been used to refer to the various frames. We will use a simple notation in which the reference frame is denoted by +0 and the overlap reading frame is given by the number of shifted nucleotides, preceded by a plus "+" or minus "-" sign indicating whether the overlap occurs in the same or opposite direction, respectively (Fig. 1). For example, overlap in the opposite direction by shifting two nucleotides to the right is denoted by reading frame -2.

The first overlapping genes were discovered in viruses (Barrell et al. 1976; Fiddes and Godson 1979), but they are also known to be present in prokaryotic (in particular bacterial) and eukaryotic genomes. The existence of overlapping genes in bacterial genomes was strongly denied initially and, consequently, putative overlapping genes were excluded from annotation programs (Warren et al. 2010). Although experimental verification of two protein-coding genes in the same DNA sequence is challenging, an increasing number of overlapping genes have been identified in prokaryotes in recent years (for example, see Fellner et al. 2013). Overlapping genes are also present in vertebrates, in particular among *de novo* genes (Makalowska et al. 2007). It recently came to light that the number of overlapping genes could be greater than expected, particularly in the virus world. Examining the gene coordinates (from GenBank) available for virus sequences, Belshaw et al. (2007) pointed out 819 instances of gene overlap among the 701 reference RNA virus genomes, with 56% of the viruses having at least one gene overlap. A recent study by Pavesi et al. (2013) assembled a dataset of 27 experimentally-identified overlapping genes longer than 140 nucleotides.

Gene overlap is thought to play a major role for increasing the coding space in genomes highly constrained for their maximum length (Miyata and Yasunaga 1978; Sakharkar et al. 2005), notably in viruses (Chirico et al. 2010). It has been shown to participate in gene regulation of functionally related proteins (Normark et al. 1983; Cooper et al. 1998; Zheng et al. 2002; Wagner et al. 2002). Gene overlap also seems to be involved in the creation of new genes, or "overprinting" (Keese and Gibbs, 1992), particularly in viruses (Pavesi et al. 2013), but also in the human genome (Knowles & McLysaght, 2009). In addition, overlapping coding regions could represent a process with the beneficial effect of limiting the evolution of these regions in viral genomes (Simon-Loriere et al. 2013). Indeed, the evolution of double-coding nucleic acid sequences is highly constrained, as any mutation affects both genes and may affect both associated proteins. Overlapping coding regions in viruses could thus represent favorable therapeutic targets, where viruses have less freedom to escape drug effects.

The first mathematical analyses of this double-coding were published in the early 1980s (Sander and Schultz 1979, Siegel and Fitch 1980; Smith and Waterman 1980). As the genetic code is degenerated (18 of the 20 amino acids are encoded by several codons, Tab. 1), it is possible to measure the information contained in the amino acid sequence encoded by a gene in the reference frame with respect to possible overlapping genes in the five other reading frames. These theoretical considerations, based in particular on information theory, have shown that the five possible overlapping reading-frame configurations differ significantly in their coding flexibility and thus in their information content. In particular, frame -2 is highly constrained and thus expected to be "very rare in nature" (Smith and Waterman 1980). However, the existence of overlapping genes in the anti-sense reading frame -2 has been suggested in several viruses. For example, in HIV-1, Miller (1988) discovered a long Open Reading Frame of ~185 codons overlapping the *env* gene in frame -2, and named it the "AntiSense Protein" (ASP) gene. The existence and function of ASP are still controversial, but several pieces of evidence argue in favor of this hypothesis (see Torresilla et al. 2015 for a review, and two recent studies showing an immune response *in vivo*: Berger et al., 2015, Bet et al., 2015). Moreover, we demonstrated (Cassan et al. 2016) that a strong selection pressure acts to maintain the ASP ORF in the group M of HIV-1, which is responsible for the human pandemic. Interestingly, out of the 819 pairs of overlapping genes pointed out in viruses by Belshaw et al. (2007), +1 frameshifts are significantly more frequent than +2 frameshifts, while these two same sense overlaps are comparable regarding information theory predictions (Smith and Waterman, 1980). Belshaw et al. (2007) proposed several possible explanations for this difference, notably related to codon usage.

Specific evolutionary models have been developed for DNA sequences encoding two overlapping genes (*e.g.* Hein and Stovlbaek, 1995; Krakauer, 2000; Pedersen and Jensen, 2001), and more recently for detecting selection pressure specific to double-coding (e.g. Sabath et al., 2008, 2011, Pavesi et al., 2013; Simon-Loriere et al., 2013; Firth, 2014; Mir and Schober, 2014, Wei and Zhang, 2015). However, little has been done to evaluate the interdependent constraints elicited by the overlap on the amino-acid sequence of the associated proteins. Some studies reported that overlapping protein regions are enriched in amino acids with high codon degeneracy (Pavesi et al., 1997) - that is, Arginine (R), Leucine (L) and Serine (S), which are encoded by six differing codons - and that they often respectively encode a cluster of either basic or acidic amino acids (Pavesi, 2000). More recently, Rancurel et al. (2009) showed a significant bias towards structural disorder, that is, the lack of a rigid 3D structure for protein regions encoded by genes overlapping in the same sense (frame +1 or +2).

In this paper, we study constraints (and their combinatorics) acting on the composition of two overlapping proteins. First, their amino-acid composition is subject to specific constraints depending on the overlapping frameshift. For example, a Tyrosine (Y) in reading frame -2 always overlaps with a Tyrosine in the reference frame, and reciprocally. More constraints appear when considering the protein composition in terms of di-peptides. In particular, some constraints reveal "forbidden" di-peptides, that is di-peptides that never occur in one frame (reference or overlap), as it would necessarily induce a Stop codon in the other frame (e.g. YY in frame -2). The biological meaning of these results is discussed, yielding support for understanding the mechanism of overlapping genes and for developing overlapping gene detection methods. After introducing notation and first results in section 2, we propose a graph traversal algorithm for deriving a set of frameshift specific constraints in terms of amino-acid composition in Section 3. Di-peptide constraints are analyzed in Section 4 and generalization to $n$-peptides is given in Section 5.

|   |   | t |   |   | c |   |   | a |   |   | g |   |
|---|---|---|---|---|---|---|---|---|---|---|---|---|
| t | ttt<br>ttc | Phe | F | tct<br>tcc<br>tca<br>tcg | Ser | S | tat<br>tac | Tyr | Y | tgt<br>tgc | Cys | C |
|   | tta<br>ttg | Leu | L |   |   |   | taa<br>tag | Stop |   | tga | Stop |   |
|   |   |   |   |   |   |   |   |   |   | tgg | Trp | W |
| c | ctt<br>ctc<br>cta<br>ctg | Leu | L | cct<br>ccc<br>cca<br>ccg | Pro | P | cat<br>cac | His | H | cgt<br>cgc<br>cga<br>cgg | Arg | R |
|   |   |   |   |   |   |   | caa<br>cag | Gln | Q |   |   |   |
| a | att<br>atc<br>ata | Ile | I | act<br>acc<br>aca<br>acg | Thr | T | aat<br>aac | Asn | N | agt<br>agc | Ser | S |
|   | atg | Met | M |   |   |   | aaa<br>aag | Lys | K | aga<br>agg | Arg | R |
| g | gtt<br>gtc<br>gta<br>gtg | Val | V | gct<br>gcc<br>gca<br>gcg | Ala | A | gat<br>gac | Asp | D | ggt<br>ggc<br>gga<br>ggg | Gly | G |
|   |   |   |   |   |   |   | gaa<br>gag | Glu | E |   |   |   |

**Table 1: Standard genetic code.**

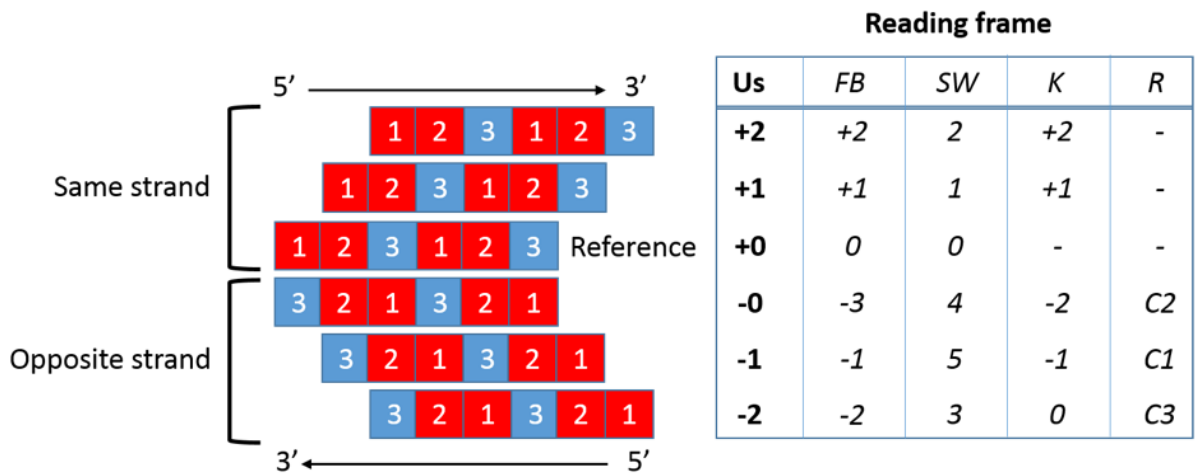

**Figure 1**: **Our notation (Us) for the six overlapping reading frames, and correspondence with alternative notations found in literature**: FB (Firth and Brown, 2006), SW (Smith and Waterman, 1980), K (Krakauer, 2000), R (Rogozin et al., 2002).

## 2. Notations and first results

*Overlapping frame reciprocity*
Reciprocity underlies the various overlapping frames and is specific to the direction of the overlap. When considering overlap in the same direction, the second reading frame (+1 or +2) depends on the sequence chosen as the reference (frame +0). If a sequence $x'$ overlaps a reference sequence $x$ with frame +1, then, in turn, sequence $x$ overlaps sequence $x'$ with frame +2. For the three types of overlap occurring in the opposite direction (-0, -1 or -2), the overlapping reading frame with respect to the reference frame is the same regardless of the sequence chosen as the reference (e.g., reading sequence $x$ in frame +0 and $x'$ in frame -1 leads to the same overlap as reading $x'$ in frame +0 and $x$ in frame -1). Although very simple to derive from observing the five overlapping reading frames (Fig. 1), these reciprocity properties are important, as they generate symmetries among the constraints induced on proteins coded by overlapping genes.

*Quadons*
Overlap in the opposite strand without shifting (i.e. frame -0) is a particular case where the two overlapping codons fully overlap: sites 1, 2, 3 in the reference frame respectively match sites 3, 2, 1 in the overlap. When considering overlap with shifting (frame +1, +2, -1 or -2, Fig. 1), the overlapping codons only have two DNA bases in common. Thus, a suite of four nucleotides is required for a complete description of two overlapping codons. A tetra-nucleotide or "quadon", taken at a chosen position in the sequence, describes exactly one codon in each reading frame. Both the starting site of a quadon describing two overlapping codons, and the reading frames of the two overlapping codons inside the quadon, are specific to the type of overlap. To simplify reading, we use an explicit notation (derived from Rogozin et al. 2002) for quadons and the corresponding codons. For example, if we assume the quadon sequence in the reference frame is **acat**, then we have the following overlapping pairs of quadons in each frame (remember that in DNA double helix, the reverse complement is defined by the nucleotide pairs a↔t and c↔g):

    Frame +1        **acat**/*acat*        Frame +2 **acat**/*acat*
    Frame -1        **acat**/*atgt*        Frame -2 **acat**/*atgt*

where bold and italic letters are used for the reference and the overlap, respectively; both codons are underlined and read from left to right.

*Fewer Stop codons in frame -2*

From the genetic code (Tab. 1), a Stop codon can be encoded by three codons **taa**, **tag**, and **tga**. Thus, for each of the two overlapping reading frames, 12 differing quadons include a Stop (three possible Stop codons times four possible nucleotides for the remaining letter of the quadon), totaling 24 quadons that are never observed in double-coding sequences. However, in frame -2 and only in that frame, two Stop codons may overlap (themselves or each other): **taa** and **tag**. This leads to four "double Stop" quadons denoted using the above-defined notation as **ttaa**/*ttaa*, **ctaa**/*ttag*, **ttag**/*ctaa*, and **ctag**/*ctag*. Therefore, only 20 quadons are not observable in frame -2. Moreover, the absence of a Stop codon in the reference frame reduces the probability of observing a Stop in the overlapping frame -2. As only half of the **ta\*** codons are permitted in the reference frame (" * " represents any nucleotide; **tat** and **tac** are permitted, whereas **taa** and **tag** are forbidden), the frequency of **taa** and **tag** Stop codons should be divided by two in frame -2, compared to other frames, assuming uniform codon usage. Consequently, the probability of long open reading frames (part of a reading frame that contains no Stop codon) is increased in frame -2, compared to other frames. Such findings are indeed observed in gene coding sequences. For example, in HXB2 (the first discovered strain of HIV, Barré-Senoussi et al. (1983), Ratner et al. (1985)), Stop codons in the non-overlapping coding regions are much less frequent in frame -2 (2.5% of the codons) than in the other frames (5.5%, 5%, 10.5%, and 7.5% in frames -1, -0, +1, and +2, respectively). An additional factor explaining these striking frequency differences could be that the presence of Stop codons is positively selected in sense frames +1 and +2, to interrupt translation quickly in case of an erroneous frameshift. Seligmann and Pollock (2004) suggested that particular codon usage biases have evolved so as to increase the frequency of Stop codons in unused ORFs and hence reduce the wastage caused by the translation of accidentally frameshifted genes (see also Cusack et al. (2011) for such results in the human genome). Hence gene overlap probability depends strongly on codon usage (Belshaw et al. 2007) in particular through the Stop codon frequency in all the reading frames. Moreover, GC-rich genomes reduce the chance for spurious Stop codons in any frame, as Stop codons are made essentially of **a** and **t** nucleotides. HXB2 strain is AT-rich (57.4% of **a** and **t**); this is an additional factor to explain the high frequencies of Stop codons.

*Amino-acid composition constraints*

Two proteins encoded by overlapping genes are subject to amino-acid composition constraints which result directly from the genetic code. For example, with frame -2, two overlapping proteins are subject to the constraint that amino acids C or W in the reference frame are necessarily overlapped by amino acids H or Q in the overlapping frame -2, and reciprocally. We denote this constraint as:

**(reference frame)**    **C + W** ⇔ *H + Q*    *(overlapping frame -2)*

where the left, bold part of the constraint corresponds to the reference frame, and the italic part on the right, to the overlapping frame -2. Figure 2 illustrates the derivation of this constraint from quadons.

Such amino-acid composition constraints can be interpreted as logical equivalences, where + stands for *"or"*: in double-coding sequences, amino acid **C** or **W** in the reference frame overlaps with either amino acid *H* or *Q* in frame -2, and reciprocally. They can also be read in terms of frequency equalities as the number of occurrences of amino acids **C** and **W** in the reference frame equals the number of occurrences of *H* and *Q* in the overlapping frame -2. Note that due to reciprocity, the relation ⇔ is symmetric in frame -2 and relation **H + Q** ⇔ *C + W* also holds.

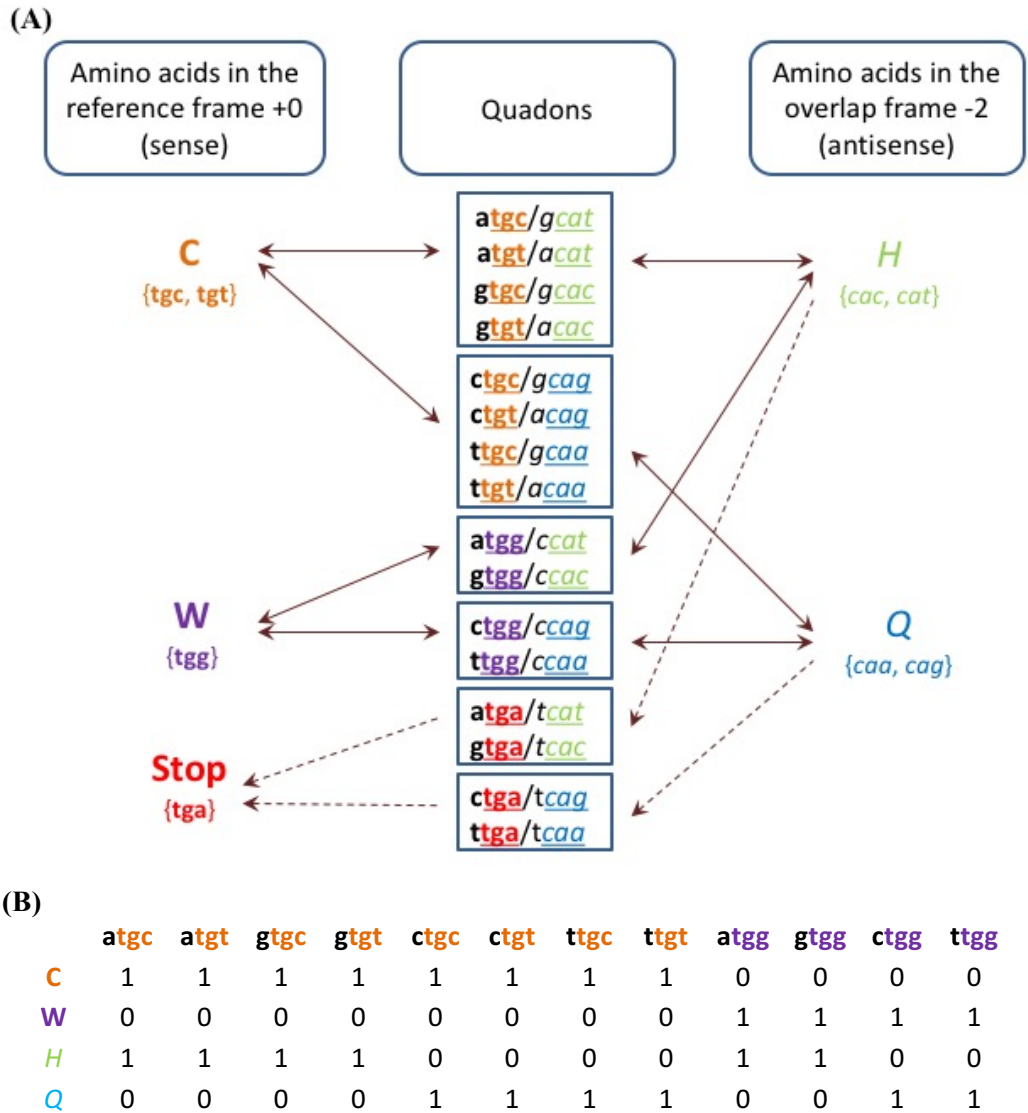

**Figure 2: Representation of amino acid constraint C + W (+Stop) ⇔ H + Q occurring in frame -2.** Bold letters (**C**, **W**, **Stop**) and italic letters (*H, Q*) correspond respectively to the amino acids read in the reference frame and the overlapping frame. **(A) Graphical representation.** The middle column lists the possible quadons (reference/frame -2) for each possible amino acid overlap. The codon read in each frame of a quadon is underlined. Amino acid *C*, which can be encoded by two codons {**tgc**, **tgt**}, is present in 8 quadons, which encode in reading frame -2 for either amino acid *H* {**cac**, **cat**} or *Q* {**caa**, **cag**}. The latter amino acid *Q* can, in turn, be overlapped by **W** {**tgg**} or by a Stop codon {**tga**}. Amino acid **W** may not be overlapped by any other amino acid and the Stop codons are assumed absent from a coding sequence. The set of amino acids involved in the constraint defines a connected component in the "tri-partite graph" associated with frame -2 (Section 3). **(B) Associated adjacency matrix.** Columns refer to the quadons (read in the reference frame), and rows to the amino acids. The rows correspond to the four amino acids (two in the reference, two in the overlap) included in the connected component of graph (A). These rows are linearly dependent, i.e. $R_1 + R_2 - R_3 - R_4 = 0$ where $R_i$ is the $i^{th}$ row of the matrix (B).

## 3. Deriving amino-acid composition constraints

In order to study the combinatorics of these logical constraints in overlapping proteins, we use a graph-based approach for listing the constraints exhaustively (not only for single peptides, but also for $n$-peptides, as described in the following sections). Codon overlap is only partial, except for the reading frame -0 (opposite overlap without shifting). This generates codon dependency along the sequence, as each amino acid read in one frame overlaps with two codons in the other frame. This aspect will become apparent when examining constraints that apply to polypeptides (Sections 4 and 5). In this section, we consider constraints that apply to single amino acids (or mono-peptides).

*Graph traversal algorithm*
The amino-acid constraints are a direct consequence of the genetic code, and they can be derived efficiently from a tri-partite graph, where the three types of vertices are: (i) the amino acids read in the reference frame, (ii) the quadons, and (iii) the amino acids read in the overlapping frame. Each quadon is linked by an edge to exactly one amino acid in the reference frame and one amino acid in the overlap. The tri-partite graph defined in this manner is frame-specific, as the two amino acids encoded by a given quadon depend on the overlapping reading frame (see the illustration with quadon **acat** in Section 2). The tri-partite graph for frame -0 is similar, but uses codons instead of quadons.

To obtain an exhaustive list of amino-acid constraints, we leverage the fact that each connected component (i.e. maximal sub-graph in which any two vertices are connected to each other by paths) in the full tri-partite graph (including the $4^4$ possible quadons) defines an irreducible equivalence constraint, binding a set of amino acids in the reference frame with a set of amino acids in the overlapping frame. For example, Figure 2 describes the connected component associated with the amino-acid constraint **C** + **W** $\Leftrightarrow$ *H* + *Q* occurring in frame -2.

We implemented a graph traversal algorithm for finding the set of all of these connected components and thus listing the corresponding irreducible amino-acid constraints. The search begins at a chosen amino acid, finds the set of amino acids included in the connected component containing this amino acid, and then returns the associated amino acid constraint. Our algorithm is outlined in Figure 3. In order to find all the connected components of the graph, we loop through the 20 amino acids in the reference frame, starting a new search whenever the loop reaches an amino acid that has not already been included in a previously-found connected component. This algorithm is similar to conventional graph traversal algorithms allowing the computation of connected components in linear time in terms of the number of edges in the graph (Hopcroft and Tarkan, 1973), that is, here, twice the number of quadons in the tri-partite graph (see Figure 2). This graph traversal algorithm is easily extended to polypeptides by replacing quadons (four nucleotides) with "heptons" (seven nucleotides) for di-peptides and, more generally, by DNA words of length $3n + 1$ for $n$-peptides. This algorithm is implemented in the R package DCODE available from the CRAN: https://cran.r-project.org/web/packages/DCODE.

Other frameworks may be used to derive these constraints. Using a linear algebraic approach, graph connectivity is described by an adjacency matrix of size 40 (total number of amino acids in both frames, reference and overlap) times the number of quadons ($4^4$). From graph theory, the adjacency matrix has rank $40 - c$, where $c$ is the number of connected components of the graph (for instance, see (Biggs 1993), p.25). Each connected component corresponds to a set of linearly dependent rows of the adjacency matrix (see Fig. 2 for an illustration with constraint **C** + **W** $\Leftrightarrow$ *H* + *Q* in frame -2). Finding the connected components comes down to finding the null linear combinations of the rows of the adjacency matrix. This approach works nicely for deriving amino-acid constraints, but requires more computations than a single graph traversal algorithm, in particular when increasing the length of the overlapping peptides.

**Input:** An amino acid $X$ and a frame $f$.
**Output:** Two sets of amino acids: $A_1$ (containing amino acid $X$) and $A_2$, respectively in the reference and in the overlapping frame $f$, bound by the amino-acid constraint $A_1 \Leftrightarrow A_2$.

1. Set $A_1 = \{X\}$, $A_2 = \emptyset$.

2. Generate $Q$, the set of quadons (4-nucleotide sequences) coding for amino acid $X$ according to the genetic code in the reference frame.

3. Generate $\overline{Q}$, the set of quadons overlapping the quadons $Q$ in frame $f$ (after shift and reverse complement if necessary).

4. Generate $\overline{A}$, the set of amino acids encoded in frame $f$ by the overlapping quadons $\overline{Q}$.

5. **if** $\overline{A} \neq \emptyset$, i.e. if the overlapping quadons in $\overline{Q}$ do not all encode for a Stop, **then**

   (a) Add the set of amino acids $\overline{A}$ to the set $A_2$.

   (b) Generate $\overline{Q}^+$, the set of <u>all the other</u> quadons (with the exception of the quadons in $\overline{Q}$) coding for the amino acids in $\overline{A}$.

   (c) **if** $\overline{Q}^+ \neq \emptyset$ **then**

       i. Generate $\overline{\overline{Q}}$, the set of quadons in the reference frame associated with the quadons in $\overline{Q}^+$.

       ii. Generate $A$, the set of amino acids encoded by the quadons in $\overline{\overline{Q}}$.

       iii. **if** $A \neq \emptyset$ **then**
   add $A$ to the set $A_1$ and generate $\overline{\overline{Q}}^+$, the set of all the other quadons (with the exception of the quadons in $\overline{\overline{Q}}$) coding for amino acids in $A$.

       **if** $\overline{\overline{Q}}^+ \neq \emptyset$ **then**
   set $Q = \overline{\overline{Q}}^+$ and go to (3).

6. Return the associated amino-acid constraint: $A_1 \Leftrightarrow A_2$.

**Figure 3: Tri-partite graph traversal algorithm.** This algorithm lists the set of amino acids (reference and overlap) defining a connected component in the tri-partite graph for a given overlapping frame in {-2, -1, -0, 1, 2}, and returns the associated amino-acid composition constraint.

| Frame | Reference | | Overlap | |
|---|---|---|---|---|
| | A | ⇔ | A | (1) |
| | Y (+ Stop) | ⇔ | Y (+ Stop) | (2) |
| | G | ⇔ | P | (3) |
| | P | ⇔ | G | (4) |
| | T | ⇔ | V | (5) |
| -2 | V | ⇔ | T | (6) |
| | H + Q | ⇔ | C + W (+ Stop) | (7) |
| | C + W (+ Stop) | ⇔ | H + Q | (8) |
| | I + M | ⇔ | I + M | (9) |
| | + Trivial constraint: | | | |
| | D + E + F + K + L + N + S + R | ⇔ | D + E + F + K + L + N + S + R | (10) |
| | F + L | ⇔ | E + K + Q (+ Stop) | (1) |
| | E + K +Q (+ Stop) | ⇔ | F + L | (2) |
| -0 | I + M + V | ⇔ | D + H + N + Y | (3) |
| | D + H + N + Y | ⇔ | I + M + V | (4) |
| | + Trivial constraint: | | | |
| | A + C + G + P + R + S + T + W (+ Stop) | ⇔ | A + C + G + P + R + S + T + W (+ Stop) | (5) |
| | D + H + N + Y | ⇔ | I + M + T | (1) |
| +1 | + Trivial constraint: | | | |
| | A + C + E + F + G + I + K + L + M + P + Q + R + S + T + V+ W (+ Stop) | ⇔ | A + C + D + E + F + G + H + K + L + N + P + Q + R + S + V + W + Y (+ Stop) | (2) |
| | I + M + T | ⇔ | D + H + N + Y | (1) |
| +2 | + Trivial constraint: | | | |
| | A + C + D + E + F + G + H + K + L + N + P + Q + R + S + V + W + Y (+Stop) | ⇔ | A + C + E + F + G + I + K + L + M + P + Q + R + S + T + V+ W (+ Stop) | (2) |
| -1 | Trivial constraint only: All amino acids (+ Stop) ⇔ All amino acids (+ Stop) | | | (1) |

**Table 2: Amino-acid constraints for the five overlapping frames.** For the sake of completeness, we also indicate the locations at which the Stop codons may occur; for example, in frame -2, a Stop codon may overlap a Y (Tyrosine). The overlap with reading frame -2 shows the greatest number (10) of constraints and includes the most specific ones, whereas frame -1 is subject to a unique, trivial constraint. The four major constraints in frame -0 oppose strongly hydrophobic amino acids (blue) in one frame to strongly hydrophilic amino acids (green) in the other frame, that is (Kyte Doolittle (1982) hydropathy index within brackets): L (3.8) and F (2.8) overlap Q (-3.5), E (-3.5) and K (-3.9) in Constraints (1, 2); whereas I (4.5), V (4.2) and M (1.9) overlap Y (-1.3), H (-3.2), D (-3.5), and N (-3.5) in Constraints (3, 4).

*Amino-acid constraints list*
The exhaustive list of the frame-specific amino-acid constraints obtained with our graph traversal algorithm is provided in Table 2. As expected, the overlap with reading frame -2 shows the greatest number of constraints (10) and includes the most specific constraints (i.e., binding the most restricted subsets of amino acids). Six constraints describe a single amino acid overlap: each Alanine (A) in the overlapping frame overlaps an Alanine in the reference frame +0 (and reciprocally); similarly, the Tyrosine (Y) amino acids necessarily overlap themselves; while amino acids Glycine (G) and Proline (P) (resp. Threonine (T) and Valine (V)) overlap each other. The latter is a strong constraint, as Glycine is frequent in standard proteins (the 3rd most frequent amino acid (Eitner et al. 2010) in proteins from the UniProt TREMBL database), whereas Proline is established as a potent breaker of both alpha-helical and beta-sheet protein structures, and is among the least frequent amino acids in proteins (the 12th most frequent). At the amino acid scale, the number of constraints for the other types of overlap is more limited: frame -0 overlap is subject to 5 amino-acid constraints; same sense overlaps (frame +1 and +2) are subject to two amino-acid constraints; and frame -1 is only subject to a unique, trivial constraint. Note that this trivial constraint holds true in all frames given that a double-coding sequence has the same number of amino acids in each frame (reference and overlap). However, in all the frames except frame -1, this trivial constraint can be reduced. Hence, the number of effective constraints within each overlapping frame is better viewed as the total number of constraints minus one (nine effective constraints in frame -2, four in frame -0, only one in frames +1 and +2, and none in frame +1). Moreover, we (unfairly) call the largest constraint of each frame the "trivial" constraint (Tab. 2).

Most of the (numerous) constraints in frame -2 are interpreted easily. The first two letters in this frame (on their own almost sufficient to determine an amino acid) of a codon in the reference also face the first two letters of the codon in the overlap. Thus, the 2nd and 3rd letters of a quadon restrict the possibilities to only one or two amino acids in both the reference and the overlap. For example, the constraint **C + W** ⇔ *H + Q* (Fig. 2) is derived from the set of quadons with form **xtgy**/*ycax*, where **x** and **y** represent any of the four nucleotides, and *x*, *y* are their reverse complements; moreover, codon **tgy** in the reference reading frame and codon *cax* in the overlap respectively code for {C, W or Stop} and {H or Q} (Tab. 1). Another simple example is **A** ⇔ *A* (Constraint (1)), which is induced by the quadons with the form **xaty**/*yatx*. Again, the set of quadons with the form **xtay**/*ytax* induces **Y(+Stop)** ⇔ *Y(+Stop)* (Constraint (2)). However, Constraint (10) includes more amino acids, due to the fact that some of them, namely L, R and S, are encoded by codons starting with two different di-nucleotides, and not only ones such as C, W, H, Q, A, Y in our previous examples. For example, both codons **cty** and **tty** may code for L (Tab. 1). Consequently, we obtain a much larger connected component in the tri-partite graph. The same holds for the other frames. In particular, in frame -1, the 2nd and 3rd codon positions face each other in both strands. Thus a larger number of amino acids corresponds to any codon with a given di-nucleotide in 2nd and 3rd positions (e.g. **xtt** may code for F, L, I, or V, depending on nucleotide **x**). Hence, we obtain a unique, global connected component in the graph, defining a unique, trivial amino-acid constraint.

Each of our constraints (Tab. 2) is a logical equivalence resulting from a set of implications. Obviously, each implication may yield information that is more specific. For example, we have 5 implications in frame -0:

$$
\begin{aligned}
\mathbf{F} &\Rightarrow E+K, \\
\mathbf{L} &\Rightarrow E + K + Q + Stop, \\
\mathbf{F + L} &\Leftarrow E, \\
\mathbf{F + L} &\Leftarrow K, \\
\mathbf{L} &\Leftarrow Q,
\end{aligned}
$$

which we summarize by the equivalence (Constraint (1) in frame -0):

$$\mathbf{F + L} \Leftrightarrow E + K + Q + Stop.$$

This constraint thus divides into implications that are more specific. The consequent set can be reduced for some antecedent sets (e.g. when considering the antecedent **F** in the reference frame, we have **F** ⇒ *E+K* and the consequent set in the overlap does not include *Q* or *Stop*). However, this division of our equivalence constraints only occurs for the amino-acid constraints in frame -0, and for the largest "trivial" equivalence constraint of each frame (Tab. 2). Indeed, the effective ("non-trivial") constraints in all frames except -0 are based on implications in which the consequent set is not reduced (e.g. Constraint (8) in frame -2 (Tab. 2, Fig. 2) results from: **C** ⇒ *H + Q*, **W** ⇒ *H + Q*, **C + W** ⇐ *H* and **C + W** ⇐ *Q*).

Compared to implications, our equivalence constraints summarize the information in terms of joint amino-acid composition of two overlapping protein regions. This offers the advantage of giving a compact, comprehensive representation of information, which can be interpreted easily. For example, an overlap in frame -0 generates specific biochemical constraints in terms of hydropathy. Indeed, constraints (1-4) in frame -0 oppose strongly hydrophobic amino acids to strongly hydrophilic amino acids (Tab. 2). These four constraints include most of the amino acids with strong hydropathic properties (e.g. L (3.8) and F (2.8) overlap Q (-3.5), E (-3.5), and K (-3.9), where the numbers in parentheses correspond to the Kyte and Doolittle (1982) hydropathy index). Consequently, the alternations of hydrophilic and hydrophobic amino acid stretches (which form the very basis of protein structures) in two proteins overlapping in frame -0 strongly depend on one another. For example, assuming a membrane protein being encoded on one strand, that is, a highly hydrophobic protein, we will see a highly hydrophilic protein on the other strand, typically a disordered protein with long, exposed stretches of amino acids.

### 4. Di-peptide constraints and dependency between successive codons

*Absence of dependency in frame -0*

Most di-peptide constraints result from the combination of two amino-acid constraints. This is obviously the case in frame -0, as the positions of two overlapping codons fully match. Thus, every di-peptide (and *n*-peptide) constraint corresponds to the combination of two (*n*) amino-acid constraints. For example, the combination of amino-acid Constraints (1) and (3) in frame -0 (Tab. 2), leads to four di-peptide constraints, depending on the ordering of the two amino-acid constraints:

$$(F + L) (F + L) \Leftrightarrow (E + K + Q) (E + K + Q)$$
$$(F + L) (I + M + V) \Leftrightarrow (D + H + N + Y) (E + K + Q)$$
$$(I + M + V) (F + L) \Leftrightarrow (E + K + Q) (D + H + N + Y)$$
$$(I + M + V) (I + M + V) \Leftrightarrow (D + H + N + Y) (D + H + N + Y)$$

As the di-peptide in the reference frame (bold letters) and the overlap (italic letters) are read in opposite directions, the combination of the two amino-acid constraints is obtained by concatenating the amino acids from the right part and from the left part of the constraints in opposite order ((**1**, Left) (**3**, Left) ⇔ (**3**, Right) (**1**, Right) and reciprocally). Moreover, the product notation denotes letter concatenation and thus the factor ordering matters. For example, (**F** + **L**) (**I** + **M** + **V**) describes the set of six di-peptides obtained by concatenating F or L (first) with I, M, or V (second). Hence, overlapping proteins in frame -0, subject to 5 amino-acid constraints (Tab. 2), are subject to $5^2$ di-peptide constraints, but no information is added in practice, compared to the 1-peptide analysis.

*Extension of the graph traversal algorithm*

In all overlapping reading frames except -0, the constraints cannot be derived in a straightforward manner from the amino-acid constraints due to the dependencies between successive overlapping codons (each codon overlaps two codons in the shifted reading frame). Clearly, the constraints resulting from the combination of two amino-acid constraints still hold, but they may be decomposed into sub-constraints (when corresponding to several, different, connected components in the tri-partite graph for di-peptides). To compute these constraints, we thus launched the extension of our graph algorithm (Fig. 3) using "heptons". The total numbers of di-peptide (and amino-acid) constraints for all frameshifts are given in Table 3. The additional constraints exist mainly because some di-peptides induce a Stop codon in the alternative reading frame systematically and thus are forbidden in an overlapping context. However, other constraints also appear in frame -2, and only in that frame. We discuss these two types of additional constraints in the following sub-sections.

*Additional "null" constraints for the di-peptides inducing Stop codons in the overlap*

In most overlapping frames, some di-peptides induce a Stop codon in the alternative reading frame systematically. Whichever codon is used for coding each of the two successive amino acids, the overlap is necessarily a Stop codon. Such di-peptides cannot be observed in a double-coding region with this frameshift, thus generating what we call "null" constraints. For example, the di-peptide YY in the reference frame necessarily generates a Stop codon in the overlap with frame -2. Indeed, whichever codon is used for coding each Y (**tat** or **tac**), the overlapping codon in frame -2 is either **taa** or **tag**, that is, a Stop codon. Then YY is never observed in a coding sequence with an overlapping gene in frame -2. Due to the reciprocity of frame -2 (Section 2), the absence of YY is also true for the overlapping frame. The set of forbidden di-peptides for each frame is given by the set of isolated di-peptides in the associated tri-partite graph (as a forbidden di-peptide is always overlapped by a Stop). Our graph traversal algorithm allows us to quickly identify the exhaustive set of forbidden di-peptides for all frames (null constraints in Tab. 3); we then find connected components with only one di-peptide on one side (reference or overlap). The reciprocity of the reference and overlap frames (Section 2) induces reciprocity of the null constraints. In total for frame -2, six di-peptides necessarily code for a Stop codon in the overlapping frame: CY, DY, FY, HY, NY, YY, which results in 12 "null" constraints, implying the absence of these di-peptides in both strands. A particularity of frame -0 is that none of the di-peptides is overlapped by a Stop codon systematically, and thus no null-constraint is found. For same sense overlaps, none of the di-peptides in the overlap with frame +1, or in the reference for frame +2, induces a Stop systematically in the other overlapping frame, and we only have five null-constraints, without reciprocity.

Di-peptides in Table 3 are special in that they impose Stop codons in alternative reading frames and are thus forbidden in an overlapping context. This notion of forbidden di-peptide is based on simple considerations that are directly from genetic code. However, to the best of our knowledge, this notion and the lists of the di-peptides forbidden in frame-specific overlapping contexts have not been suggested before in the large body of literature dedicated to overlapping genes. In addition, several authors (e.g. Cusack et al. (2011)) have shown that Stop codons are more frequent than expected in alternative reading frames, in order to limit the translation of accidentally frameshifted transcripts. We thus expect that these forbidden di-peptides could be favored in regions that are free of gene overlap.

| Frame | Mono-peptide | Di-peptide | | | Forbidden di-peptides | |
|---|---|---|---|---|---|---|
| | | All | Non-null | Null | Reference | Overlap |
| -2 | 10 | 125 | $10^2 + \mathbf{13}$ | **12** | CY, DY, FY, HY, NY, YY. | CY, DY, FY, HY, NY, YY. |
| -1 | 1 | 11 | 1 | **10** | FI, FK, FM, FN, FT. | FI, FK, FM, FN, FT. |
| -0 | 5 | 25 | $5^2$ | **0** | ∅ | ∅ |
| +1 | 2 | 9 | $2^2$ | **5** | MI, MK, MM, MN, MT. | ∅ |
| +2 | 2 | 9 | $2^2$ | **5** | ∅ | MI, MK, MM, MN, MT. |

**Table 3: Mono- and di-peptide constraints for each overlapping frame.** We provide the number of constraints and list of the "forbidden" di-peptides (inducing a Stop codon in the overlapping frame). For each frame, except frame -0 where the codons fully overlap, the number of di-peptide constraints is greater than the square of the number of amino-acid constraints. The bold numbers refer to the number of additional di-peptide constraints that cannot be derived from the amino-acid constraints: additional "non-null" constraints for frame -2 only, and additional "null" constraints for all frames except frame -0. The last two columns list the forbidden di-peptides corresponding to "null" constraints in the reference and overlap frames.

*Additional "non-null" constraints in frame -2 only*
Based on the results of our graph algorithm (Tab. 3), it turns out that for all overlapping frames except frame -2, the set of di-peptide constraints is given exactly by the ordered combination of two amino-acid constraints, plus the null constraints generated by the "forbidden" di-peptides. In frame -2, the codon dependency also generates novel "non-null" di-peptide constraints, which cannot be derived from the amino-acid constraints. For example, the constraint **SY** ⇔ *Y (D+E)* existing in frame -2 (see Appendix), cannot be deduced from constraints for amino acid S and Y (constraints (10) and (2) in Tab. 2) as amino acid S is included in the large constraint (10) involving eight amino acids in each frame.

Actually, the ordered combination of constraints (10) and (2) in frame -2, leads to two di-peptide constraints:

$$SY \Leftrightarrow Y (D+E)$$

$$(E+K+L+R) Y \Leftrightarrow Y (F+K+L+N+R+S)$$

while the simple ordered combination would produce one single constraint:

$$(D + E + F + K + L + N + S + R) Y \Leftrightarrow Y (D + E + F + K + L + N + S + R)$$

Clearly, this latter constraint is valid, but it is subdivided into the two preceding constraints, which have been simplified by accounting for null constraints with di-peptides DY, FY, and NY (Tab. 2). The complete list of the additional "non-null" constraints in frame -2 (which cannot be derived from the amino-acid constraints) is given in Appendix A.

This decomposition was expected for frames where the codons do not fully overlap, that is, all frames except -0. Indeed, the amino-acid constraints are derived from the quadons where only two nucleotides (sites 2 and 3 of the quadon) are constrained, whereas the nucleotides at both ends of the quadon (sites 1 and 4) are kept free as they are involved in only one codon (either in the reference or in the overlap). When considering di-peptides, the constraints are induced by the overlap of two codons, that is, "heptons", where five nucleotides (sites 2 to 6) are constrained, instead of four (2x2) with the two corresponding quadons. Thus di-peptide overlap would be expected to generate more specific constraints than amino-acid overlap. Surprisingly, our results (Tab. 3) show that this expectation occurs only in frame -2. In the other overlapping frames with shifting {-1, +1, +2}, the "non-null" di-peptide constraints simply correspond to the ordered combinations of two amino-acid constraints. Thus, the analysis of the non-null di-peptide constraints in frames {-1, +1, +2} brings no practical contribution as they can be deduced directly from the amino-acid constraints. Analysis of the non-null di-peptide constraints only matters for frame -2, where the dependence structure brings up rules at level 2 (di-peptides) which are not deduced from level 1 (amino acids). In the next section, we shall see that the same is found with $n$-peptides, $n > 2$.

## 5. What about higher length peptides?

***Counting the $n$-peptide null constraints***

For all frames, the peptides of length $n$ inducing a Stop codon in the alternative coding sequence are simply those containing a "forbidden" di-peptide (Tab. 3). Hence, the number of null constraints for $n$-peptides is computed recursively from the list of forbidden di-peptides for each frame. Using the recursive formula, we easily derive a closed formula to count the null constraints associated with that frame. Again, we will detail these calculations with frame -2, but the approach is similar for the other frames (detailed in Appendix B).

Let us denote by $A_n$ the set of "allowed" peptides of length $n$, that is, those not containing any forbidden di-peptide in the reference frame. Due to reciprocity, this set of allowed $n$-peptides is the same in the overlap frame. The set of di-peptides inducing a Stop codon in the overlapping frame -2 is {CY, DY, FY, HY, NY, YY} (Tab. 3). The set $A_n$ can be divided into two disjoint subsets:

$$A_n = A_{n,Y} \cup A_{n,not\{Y\}}$$

where $A_{n,Y}$ is the subset containing the allowed $n$-peptides ending with amino acid Y, and $A_{n,not\{Y\}}$ is the subset made of the allowed $n$-peptides ending with any amino acid except Y. The subset $A_{n,not\{Y\}}$ contains exactly the set of allowed $(n-1)$-peptides suffixed with an amino acid different from Y, as suffixing one of the $(20-1)$ amino acids different from Y will not generate a forbidden di-peptide (necessarily ending with amino acid Y). Thus, $A_{n,not\{Y\}}$ has size $(20-1)|A_{n-1}|$, where $|.|$ denotes cardinality. The subset $A_{n,Y}$ contains the elements of the subset $A_{n-1,not\{C,D,F,H,N,Y\}}$ of allowed $(n-1)$-peptides ending with any amino acid except {C, D, F, H, N, Y}, suffixed with amino acid Y. In turn, $A_{n-1,not\{C,D,F,H,N,Y\}}$ is made of the allowed $(n-2)$-peptides suffixed with any amino acid except {C, D, F, H, N, Y} and thus sizes $(20-6)|A_{n-2}|$. Then the total number of allowed $n$-peptides in frame -2 (reference or overlap) is computed recursively as follows:

$$\forall n > 2, \ |A_n| = |A_{n,not\{Y\}}| + |A_{n,Y}| = 19|A_{n-1}| + 14|A_{n-2}|$$
$$|A_1| = 20 \ \text{and} \ |A_2| = 20^2 - 6 = 394$$

This equation defines a second order linear homogeneous recurrence relation with constant coefficients $|A_n| = c_1|A_{n-1}| + c_2|A_{n-2}|$, which is satisfied for all $n > 2$. The same coefficients yield the characteristic polynomial, $p(t) = t^2 - c_1 t - c_2$, with roots $r_1$ and $r_2$ (if distinct). The solution of the recurrence is given by:

$$\forall n > 2, \ |A_n| = k_1 r_1^n + k_2 r_2^n$$

where coefficients $k_1$, $k_2$ fit the initial conditions of the recurrence (Greene and Knuth, 1982, section 2.1.1). This leads to the following expression of the roots $r_1$, $r_2$, and coefficients $k_1$, $k_2$:

$$r_1 = \tfrac{1}{2}\left(19 + \sqrt{19^2 + 56}\right) \approx 19.710; \quad r_2 = \tfrac{1}{2}\left(19 - \sqrt{19^2 + 56}\right) \approx -0.710;$$
$$k_1 = \tfrac{|A_2| - r_2|A_1|}{r_1(r_1 - r_2)} \approx 1.014; \quad k_2 = \tfrac{|A_1| - k_1 r_1}{r_2} \approx -0.014;$$

The total number of null constraints in frame -2 is thus equal to $2\times(20^n - |A_n|)$, where factor 2 accounts for the two frames (reference and overlap). Given that the absolute value of $r_2$ is lower than 1, the number $|A_n|$ of allowed $n$-peptides in each frame is approximated for large values of $n$ by $|A_n| \sim r_1^n$, and consequently the number of forbidden $n$-peptides by $2(20^n - r_1^n)$.

Similar computations yield recursive and closed formulae for the number of null constraints in the other frames, which are summarized in Table 4 (see Appendix B for details). Note that the total number of null constraints (including reference and overlap) equals $2\times(20^n - |A_n|)$ for anti-sense overlap (frames -2 and -1), whereas it equals $(2\times 20^n - |A_n|)$ for same sense overlap (frames +1 and +2), as forbidden peptides appear in only one of the two alternative frames (Tab. 3).

A consequence of these results is that the fraction of allowed $n$-peptides converges to 0 when $n$ grows to infinity. This result holds true for the reference frame with frame +1, the overlapping frame with frame +2, and for both frames with frames -2 and -1. This finding is interpreted easily: the longer the sequence, the more likely it is that it will contain a forbidden di-peptide. Moreover, the probability of finding an allowed/forbidden $n$-peptide for a given frame is computed easily using our formulae. This may form the basis for a quick and simple method to detect overlapping regions within proteins. For example, assume a window of size $n = 200$, and consider frame -2: if this window contains a forbidden di-peptide {CY, DY, FY, HY, NY, YY}, then this is not an overlapping region. Oppositely, if the window does not contain any forbidden di-peptides, the probability (under a simple model, with uniform amino-acid frequencies) of observing such a 200-peptide is of $\frac{|A_{200}|}{20^{200}} \approx \left(\frac{r_1}{20}\right)^{200} \approx 0.05$, thus providing significant support that the window does correspond to an overlapping region in frame -2. This approximation could be refined by incorporating realistic amino-acid frequencies and/or codon usage statistics in probability computations. This method could be used as a first filtering step, for example to scan the very large existing protein databases quickly (e.g. The UniProt Consortium, 2015). Clearly, another option involves using the genes encoding these proteins directly, and checking for the absence of Stop codons at the DNA level in the alternative reading frames. At this point in time, we can hardly envisage a situation where one would have a long peptide sequence and not have the corresponding nucleotide sequence, and thus the search for Stop codons and long ORFs remains the simplest approach for detecting overlapping genes. However, proteomic sequencing (mass spectrometry or Edman degradation) is still used, in particular for exploring variants, and could become popular again with novel technological advances, thus making this simple method useful for a first check.

### *Counting the n-peptide non-null constraints*

Due to the codon independency in frame -0 (see Section 4), the non-null, $n$-peptide constraints are derived easily in this frame, and their number equals $5^n$. For all the other frames, we computed the non-null constraints for $n$-peptides up to length $n = 5$ with our graph traversal algorithm extended to $n$-peptides (Section 3). As expected from the di-peptide analysis (Section 4), the number of non-null $n$-peptide constraints equals $2^n$ in frame +1 and +2, 1 in frame -1, and is larger than $10^n$ in frame -2 (Tab. 4). We did not achieve formal proof of these results for frame +1/+2, or find an exact formula for frame -2. All of our results are displayed in Table 4. A complete list of the non-null $n$-peptide ($n \leq 5$) constraints in frame -2 is available at: http://www.univ-montp3.fr/miap/~lebre/DCODE.html.

| Constraint type | Frame | 1-peptide | 2-peptide | 3-peptide | 4-peptide | 5-peptide | $n$-peptide |
|---|---|---|---|---|---|---|---|
| Null | -2 | 0 | 12 | 468 | 13860 | 365892 | $2(20^n - 19.710^n)$ |
|  | -1 | 0 | 10 | 400 | 11950 | 317000 | $2(20^n - 19.747^n)$ |
|  | 0 | 0 | 0 | 0 | 0 | 0 | 0 |
|  | +1/+2 | 0 | 5 | 195 | 5780 | 152745 | $2.20^n - 19.759^n$ |
| Non-null | -2 | 10 | 113 | 1316 | 15398 | 179698 | $> 10^n$ |
|  | -1 | 1 | 1 | 1 | 1 | 1 | 1 |
|  | -0 | 5 | $5^2$ | $5^3$ | $5^4$ | $5^5$ | $5^n$ |
|  | +1/+2 | 2 | $2^2$ | $2^3$ | $2^4$ | $2^5$ | $2^n$ |

**Table 4: Total number of constraints (null and non-null) as a function of the size of the peptides for all frames.** The number of constraints for peptides with size $n$ lower than five are obtained with our graph algorithm (Sections 3 and 4). The numbers of null $n$-peptide constraints are obtained from the recursive formulae described in Section 5 and Appendix B. The number of non-null $n$-peptides constraints in frame +1/+2 and -1, with $n > 5$, are conjectured from the results with $n \leq 5$. With frame -2 we only know that this number is larger than $10^n$ (number of ordered combinations of amino-acid constraints).

## 6. Conclusion and perspectives

A formal framework and a graph algorithm allowed us to introduce and characterize the constraints that bind the polypeptide composition of two proteins encoded by overlapping genes. Each constraint is a logical equivalence resulting from a set of implications. Even though implications may give more specific information, the equivalence constraints offer the advantage of yielding compact and comprehensive information, which can be interpreted directly in terms of amino-acid and polypeptide composition constraints imposed on two overlapping proteins. For example, in frame -0, four out of the five constraints correspond to specific biochemical constraints in terms of hydropathy (Section 3). Moreover, most of the theoretical approaches for understanding the constraints imposed by overlapping genes assume that the sequence of one of the two proteins is fixed (in the reference frame), and then evaluate the remaining degrees of freedom for the overlapping protein (for example, see Smith and Waterman (1980) or, more recently, Mir and Schober (2014)). In other words, they provide directional results, typically based on logical implications or conditional probabilities. However, the sequence of a protein is never entirely fixed. All proteins have variants and (generally) a large number of degrees of freedom. Thus, a major interest of our equivalence-based approach is that our constraints are symmetrical and enable characterization of the joint composition of overlapping proteins.

In agreement with previous results derived from information theory, frame -2 (anti-sense overlap when the first two sites of the codons in the two reading frames overlap) is the most constrained among the five possible reading frames. However, considering polypeptides and not single amino acids, that is, taking into account the dependency between successive codons, highlights a second type of constraint: null constraints, stating that a polypeptide cannot be observed in a given reading frame as it would induce a Stop codon in the overlap. Except in frame -0 which is not subject to dependency, these null constraints are induced in all reading frames by a specific subset of di-peptides (six di-peptides in frame -2; five di-peptides in frames -1, +1, and +2). A logical approach allowed us to study the combinatorics of both the null and non-null $n$-peptide constraints, thus pointing out that: (i) except for frame -2, non-null constraints are deduced from the amino-acid constraints, and (ii) null constraints are deduced from the di-peptide constraints. Moreover, we give a closed formula for the number of null constraints for peptides of length $n$ in each frame, which may form the basis for a fast scanning method to select possible overlapping regions within protein sequences.

From that point on, the various frames can be studied and compared in the light of these two types of constraints. Frame -2 (the most constrained) is comparable to frame -1 in terms of null constraints. The two possible overlaps in the same sense (frame +1 or +2) are subject to null constraints in only one of the two reading frames. Non-null polypeptide constraints are derived simply using ordered combinations of amino-acid constraints, except for frame -2, where novel constraints appear when increasing the length of the overlapping peptides, due to codon dependency. Analytical computation of the exact number of non-null constraints in frame -2 is left open, and is an interesting direction for further research, to fully understand the combinatorics of double-coding in this reading frame. A general open question in this article is that of fully understanding what are the patterns and trajectories of joint evolution that can be taken simultaneously by two proteins encoded on the same portion of DNA.

## 7. Acknowledgements

We would like to thank Elodie Cassan, Antoine Gross, and Jean-Michel Mesnard for stimulating and fruitful discussions, Alain Jean-Marie for its relevant indications and references, Etienne Simon-Lorière for his helpful comments on this paper, and the anonymous reviewers for their constructive advice.

# 8. APPENDIX

## A. List of additional non-null constraints in frame -2

- The combination of constraints (7) and (8) (Tab. 2) in this order gives the 2 following di-peptide constraints:

$$C\ (H+Q) \Leftrightarrow W\ (H+Q)$$
$$W\ (H+Q) \Leftrightarrow C\ (H+Q)$$

- The combination of constraints (8) and (7) gives:

$$H\ (C+W) \Leftrightarrow Q\ (C+W)$$
$$Q\ (C+W) \Leftrightarrow H\ (C+W)$$

- The double combination of constraint (7) gives:

$$W\ (C+W) \Leftrightarrow H\ (H+Q)$$
$$C\ (C+W) \Leftrightarrow Q\ (H+Q)$$

and the double combination of constraint (8) gives the reciprocal constraints.

- The combination of constraints (2) and (10) in this order gives:

$$Y\ (D+E) \Leftrightarrow SY$$
$$Y\ (F+K+L+N+R+S) \Leftrightarrow (E+K+L+R)\ Y$$

where terms **DY**, **FY**, and **NY** are removed from the right part of the constraints as they are forbidden di-peptides (Tab. 3), *i.e.* DY = 0, FY = 0, NY = 0). The (ordered) combination of Constraints (10) and (2) gives the reciprocal constraints.

- The combination of constraints (8) and (10) in this order gives (the reverse order gives the reciprocal constraints):

$$Q\ (N+K) \Leftrightarrow F\ (C+W)$$
$$(H+Q)\ (D+E) + Q\ (F+L) \Leftrightarrow (S+N)\ (C+W)$$
$$H\ (F+K+L+N+R+S) + Q\ (S+R) \Leftrightarrow (D+E+K+L+R)\ (C+W)$$

- The combination of constraints (7) and (10) in this order gives (the reverse order gives the reciprocal constraints):

$$C\ (D+E) + W\ (D+E+F+L) \Leftrightarrow (S+N)\ (H+Q)$$
$$C\ (F+K+L+N+R+S) + W\ (S+R) \Leftrightarrow (D+E+K+L+R)\ (H+Q)$$
$$W\ (N+K) \Leftrightarrow F\ (H+Q)$$

Hence, six amino-acid constraint combinations ((2)(10), (7)(7), (7)(8), (8)(7), (8)(8), (10)(2)) lead to a di-peptide constraint that divides into two di-peptide constraints; and four combinations ((7)(10), (8)(10), (10)(7), (10)(8)) lead to three di-peptide constraints. This increases the total count of "non-null" constraints by 14. Moreover, one of the $10^2$ constraints derived from the amino-acid constraints, **YY** ⇔ *YY*, is removed from the constraint set, as it is redundant with the two null constraints excluding di-peptide YY both in the reference and in the overlapping frame (Tab. 3). This results in $10^2$ **- 1 + 14** = 113 non-null di-peptide constraints in frame -2, as found by our graph algorithm (Fig. 3) applied to "heptons".

## B. Counting the $n$-peptide null constraints in frames -1, +1, +2

In frame -1, the set of di-peptides inducing a Stop codon in the overlapping frame is {FI, FM, FT, FN, FK} (Tab. 3) and the total number of allowed $n$-peptides (reference or overlap) is:

$$\forall n > 2, \ |A_n| = |A_{n,F}| + |A_{n,not\{F\}}|$$

where $A_{n,F}$ (respectively $A_{n,not\{F\}}$) is the subset containing the allowed $n$-peptides <u>starting</u> with amino acid F (resp. starting with any amino acid except F). From $|A_{n,F}| = |A_{n-1,not\{I,M,T,N,K\}}| = |A_{n-1}| - 5|A_{n-2}|$ and $|A_{n,not\{F\}}| = 19|A_{n-1}|$ follows:

$$\forall n > 2, \ |A_n| = 20|A_{n-1}| - 5|A_{n-2}|$$
$$|A_1| = 20$$
$$|A_2| = 20^2 - 5 = 395$$

In frame +1 or +2, the set of di-peptides inducing a Stop codon in the overlapping frame is {MI, MK, MM, MN, MT} either in the reference (frame +1) or in the overlap (frame +2) (Tab. 3). The total number of allowed $n$-peptides (reference in frame +1 or overlap +2) is:

$$\forall n > 2, \ |A_n| = |A_{n,not\{M\}}| + |A_{n,M}|$$

where $A_{n,M}$ (resp. $A_{n,not\{M\}}$) is the subset containing the allowed $n$-peptides <u>starting</u> with amino acid M (respectively starting with any amino acid except F). From $|A_{n,not\{M\}}| = (20-1)|A_{n-1}|$ and $|A_{n,M}| = |A_{n-1,not\{I,K,M,N,T\}}| = (20-5)|A_{n-2}|$ follows,

$$\forall n > 2, \ |A_n| = 19|A_{n-1}| + 15|A_{n-2}|$$
$$|A_1| = 20$$
$$|A_2| = 20^2 - 5 = 395$$

Lastly, the total number of allowed $n$-peptides (reference or overlap) is:

$$\forall n > 2, |A_n| = k_1 r_1^n + k_2 r_2^n$$

where

$$k_2 = \frac{|A_1| - k_1 r_1}{r_2}$$

$$k_1 = \frac{|A_2| - r_2|A_1|}{r_1(r_1 - r_2)}$$

$$r_{1,2} = \frac{1}{2}\left(c_1 \pm \sqrt{c_1^2 + 4c_2}\right)$$

and (approximate) values are given in Table B1 for each frame.

| Frame | $c_1$ | $c_2$ | $r_1$ | $r_2$ | $k_1$ | $k_2$ |
|---|---|---|---|---|---|---|
| -1 | 20 | -5 | 19.747 | 0.253 | 1.013 | -0.013 |
| +1 (reference) +2 (overlap) | 19 | 15 | 19.759 | -0.759 | 1.012 | -0.012 |

**Table B1: Coefficients defining the closed formula for the number |A$_n$| of allowed $n$-peptides in frames -1, +1, and +2** (values of $r_1$, $r_2$, $k_1$, and $k_2$, are approximated).